\documentstyle[aps,psfig]{revtex}
\begin{document}
\draft
\title{
\hfill{\small {\bf MKPH-T-00-08}}\\
{\bf Analysis of the low-energy $\eta$NN-dynamics
within a three-body formalism}
\footnote{Supported by the Deutsche Forschungsgemeinschaft (SFB 443)}}
\author{A. Fix$^a$ and H. Arenh\"ovel$^b$}
\address{$^a$Tomsk Polytechnic University, 634034 Tomsk, Russia}
\address{$^b$Institut f\"ur Kernphysik,
Johannes Gutenberg-Universit\"at Mainz, D-55099 Mainz, Germany}
\date{\today}
\maketitle

\begin{abstract}
The interaction of an $\eta$-meson with two nucleons is studied within
a three-body approach. The major features of the $\eta NN$-system in
the low-energy region are accounted for by using a $s$-wave separable
ansatz for the two-body $\eta N$- and $NN$-amplitudes.
The calculation is confined to the $(J^\pi;T)=(0^-;1)$ and $(1^-;0)$
configurations which are assumed to be the most promising candidates
for virtual or resonant $\eta NN$-states.  The eigenvalue three-body
equation is continued analytically into the nonphysical sheets by
contour deformation.  The position of the poles of the
three-body scattering matrix as a function of the $\eta N$-interaction
strength is investigated. The corresponding trajectory, starting on
the physical sheet, moves around the $\eta NN$ three-body threshold
and continues away from the physical area giving rise to virtual
$\eta NN$-states. The search for poles on the
nonphysical sheets adjacent directly to the upper rim of the real
energy axis gives a negative result. Thus no low-lying $s$-wave
$\eta NN$-resonances were found.  The possible influence of
virtual poles on the low-energy $\eta NN$-scattering is discussed.

\end{abstract}
\pacs{PACS numbers:  13.60.Le, 21.45.+v, 25.20.Lj}

\section{Introduction}
The low-energy interaction of $\eta$-mesons with few-nucleon systems
has recently become the subject of vigorous investigations. A great
deal of attention has been devoted to the search for possible bound or
resonance states in these systems.
After an experimental study at Brookhaven \cite{Brookhaven}, which casts
some doubt upon the possibility of observing $\eta$-nuclei with
relatively large mass numbers A$\geq$12, the attention  was redirected
to the interaction of $\eta$-mesons with light nuclei
\cite{Ueda,Tryas,Shev,Garc,Rakit,Wycech}.
Recent theoretical investigations have mainly
been stimulated by the precise measurements of the photoproduction
and hadronic reactions with $\eta$-mesons in final states. The
obtained results are often interpreted as strong experimental hints
that correlated states for light $\eta$-nuclear systems might exist (see e.g.\
\cite{Wilk}).  The case in point is the strong energy dependence of the
experimental cross sections observed near the $\eta$-production threshold
\cite{Metag,np}.

In the present work we deal primarily with the
dynamical properties of the $\eta NN$-system.
The understanding of these properties is important for the following
reasons.
Firstly, the $\eta NN$-system is interesting in itself. It is the
simplest $\eta$-nuclear system which admits an exact solution within
the three-body formalism. Furthermore, it may be regarded as an
example of a three-body system in which the pairwise driving forces are
attractive and which may in principle develop a bound (or virtual)
state, or a low-energy three-particle resonance near zero energy.
Secondly, the study of
$\eta NN$-scattering may serve as a promising tool for the investigation
of the corresponding processes in complex nuclei.
The possible existence of such
correlated objects would point to the crucial significance of
two-nucleon mechanisms in the $\eta$-interaction with nuclei. This
fact in turn would require the revision of the simple first order
approximation to the $\eta$-nuclear optical potential adopted by the
currently available models \cite{BeTa,FiTr,Abu}.

The main features of the $\eta NN$-dynamics in a quasideuteron state
of spin, parity and isospin $(J^\pi; T)=(1^-; 0)$ are already
elucidated in the literature. Ueda was the first to carry out an
extensive calculation of $\eta d$-scattering within a
three-body approach \cite{Ueda} and found that the
$\eta d$-system can form a quasi-bound state with a mass of 2430 MeV
and a rather small width of about 10-20 MeV. Other calculations,
also looking for bound or resonant $\eta d$-states within the Faddeev
theory, have recently been reported by Shevchenko et al.~\cite{Shev}
as well as by Garcilazo and Pe\~na \cite{Garc}. The authors
of Ref.\cite{Shev}
confirmed qualitatively the results of \cite{Ueda} and fixed the
values of the $\eta N$-scattering length for which the previously bound
$\eta d$-state becomes a low-lying $s$-wave resonance. In Ref.\cite{Garc}
the possible existence of $\eta d$ bound states within the different
$\eta N$- and $NN$-interaction models is also studied. As for the
experimental investigations, we would like to mention the measurements
presented in \cite{np} for the reaction $np\to\eta d$ where a
visible increase of the $\eta$-meson yield over the mere phase-space
calculation was observed near threshold. Recently, Metag et
al.~\cite{Metag} have reported new results for the
cross section $\gamma d\to\eta X$. These authors also note
a strong enhancement of events in the energy region of a few MeV above
threshold. Such a feature cannot be explained within the
truncated multiple scattering approach developed in~\cite{FiAr},
where only single $\eta N$- and $NN$-rescatterings
were included in the calculation as the most important corrections
to the impulse approximation.
This discrepancy is not surprising, since the validity of the latter
approach is associated with the short range
nature of the $\eta N$- and $NN$-interactions in relation to the
characteristic internucleon distance in the deuteron. In this situation,
when thinking about the whole picture of the $\eta NN$ dynamics, one may
assume that, when the attractive forces pull all particles
together so that the two-body potentials overlap,
qualitatively new features of the resulting $\eta NN$-interaction may
be expected.

In this paper we will study the question whether the dynamical
properties of the $\eta NN$-system allow the
existence of a bound (or virtual) state or a three-body resonances
in the low-energy region.
Although this question has already been covered partially in
the above mentioned works \cite{Ueda,Shev,Garc} we would like to
reinvestigate it using
a fundamentally different method, based on a search for poles of
the scattering matrix in the complex energy plane. It
possesses several important advantages over the conventional procedure
of solving the on-shell Faddeev equation.  Besides being more
convenient for determining the exact position of $S$-matrix
poles, the method provides additional insights into the source
of appearance of these poles close to the physical domain.
Furthermore, this approach allows us to investigate
the $(J^\pi;T)=(0^-; 1)$ state which is not touched upon
in the literature to the best of our knowledge. As was established in
\cite{FiAr}, this configuration plays the dominant role in the final
state of the $\gamma d\to\eta X$ reaction near threshold.
Therefore, if one attributes the anomalous behaviour of the
$\gamma d\to\eta X$ cross section reported in \cite{Metag} to
a near lying $S$-matrix pole, one has to search it in the $(J^\pi;
T)=(0^-; 1)$ state. Broadly speaking, in view of the exclusive
character of the measurements \cite{Metag}, the marked enhancement of the
near threshold yield can be assigned to the $\eta d$ final state. But
the simple estimation shows that in order to explain the observed
results one would need an enhancement factor of about 30
what in our opinion seems unlikely. In this context, large attention
is focussed on the $(J^\pi;T)=(0^-;1)$ configuration in this paper.

Our principal tool is the three-body approach
realized  within the Alt-Grassberger-Sandhas formalism \cite{AGS}.
Since the search for the $S$-matrix
poles requires the analytical continuation of the dynamical equation
into the nonphysical sheets of the Riemann surface, the analytical
form for the driving two-body interactions is needed. For this
reason, we use a simple separable potential of rank one and
restrict our consideration to only $s$-waves in the two-body
subsystems. This approach is well justified by the
empirically established $s$-wave dominance of the low-energy $\eta N$- and
$NN$- scattering. There exist strong experimental and theoretical
evidences that the low-energy $\eta N$-interaction is dominated by
the formation of the $S_{11}$(1535) resonance. Analogously, the
$^1S_0$ and $^3S_1$ poles determine to a large extent the
nucleon-nucleon low-energy interaction.
Thus the $s$-wave isobars in the $\eta N$ and $NN$ two-body channels
are expected to be the main source of the $\eta NN$-forces. Therefore
we hope that our separable model, though being quite simple, will
reproduce the major features of the $\eta NN$-dynamics.

In Sect.~\ref{sect1} we briefly describe the three-body scattering formalism
pertinent to the present problem. Some details connected with
the driving two-body interactions as well as the main
calculational formulas are given in Sect.~\ref{sect2}. In Sect.~\ref{sect3}
the procedure of analytical continuation of the scattering
equation into nonphysical energy sheets is
described and the strategy for the search of the $S$-matrix poles is
presented. The discussion of our main results and the conclusions are
presented in the last two sections.

\section{General formalism}\label{sect1}

In this section we will briefly review the properties of the three-body
equation which will concern us in the remainder of this paper. Our starting
point is the conventional three-body scattering theory in AGS form
\cite{AGS}. The three channels comprising two interacting particles and
one spectator are labeled according to the number of the spectator.
As already mentioned in the introduction, we take
for each channel as the off-shell two-body $t$-matrix a rank-one ansatz
\begin{equation}\label{eqT}
t_k(E)=|f_k\rangle\tau_k(E)\langle f_k|\,,
\end{equation}
and we will call such an interacting pair in the following ``isobar''.
Here $|f_k\rangle$ is the vertex function, and the isobar propagator $\tau_k$
is defined as
\begin{equation}\label{eqTau}
\tau_k(E)=\frac{\gamma_k}{1-\gamma_k\langle f_k|G_k^{(2)}(E)|f_k\rangle}\,,
\end{equation}
where $G_k^{(2)}(E)$ denotes the free two-body Green's function with $E$
as total c.m.\ energy of the two-particle system.
The separable form (\ref{eqT})
enables one to reduce the three-body problem to a set of coupled effective
two-body equations for the transition amplitudes $X_{ij}$
\begin{equation}\label{eqXij}
X_{ij}(W)=(1-\delta_{ij})Z_{ij}(W)+
\sum\limits_{k=1}^3(1-\delta_{ik})Z_{ik}(W)\tau_k(E_k)X_{kj}(W)\,,
\quad (i,j=1,2,3)\,,
\end{equation}
where $W$ denotes the total three-body energy while the isobar propagator
$\tau_k$ depends explicitly on the invariant energy $E_k$ of the two-body
subsystem, which is a function of $W$ and the kinetic energy of the spectator.
The energy dependent driving terms are
\begin{equation}\label{eqZ}
Z_{ij}(W)=\langle f_i|G(W)|f_j\rangle\,,
\end{equation}
with $G(W)$ being the free
three-body Green's function. When evaluating equation (\ref{eqZ}) in
momentum space and making a partial wave decomposition, it becomes a
set of one-dimensional integral equations of Fredholm type. It is
known that for an
inhomogeneous Fredholm equation to be singular, in which case the
transition matrix $X_{ij}$ has a pole, it is necessary and
sufficient for the corresponding homogeneous equation
\begin{equation}\label{eqFij}
F_{ij}(W)=\sum\limits_{k=1}^3(1-\delta_{ik})Z_{ik}(W)\tau_k(E_k)
F_{kj}(W)\,, \quad (i,j=1,2,3)\,
\end{equation}
to have a
nonzero solution for the same value of the parameter $W$. Thus the problem
of searching for poles of the scattering matrix reduces to that
of finding those values $W$ for which the Fredholm determinant $D(W)$
of (\ref{eqXij}) vanishes
\begin{equation}\label{eqD}
D(W)=0\,.
\end{equation}

\begin{figure}
\centerline{\psfig{figure=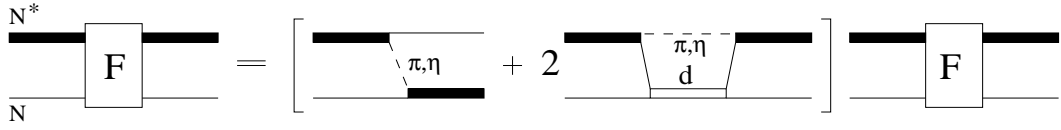,width=14cm,angle=0}}
\vspace{.5cm}
\caption{
Graphical representation of the homogeneous equation
(\protect{\ref{eqFNN}}) for the $\eta
NN$ three-body eigenvalue problem.
}
\label{fig1}
\end{figure}

We now turn to the $\eta NN$-system for which we adopt the labeling
$k=1,2$ for a nucleon spectator and $k=3$ for the $\eta$ spectator,
and furthermore change the channel notation as
\begin{equation}
\begin{array}{cl}
 1 = 2  \rightarrow  N^* &\mbox{ for the }\eta N\mbox{-isobar plus
spectator nucleon}\,, \cr
 3 \rightarrow  d &\mbox{ for the $NN$-isobar plus spectator meson}.\cr
\end{array}
\end{equation}
The identity of the channels 1 and 2 reduces the
3$\times$3 system (\ref{eqFij}) to a 2$\times$2 system which has the
following operator form in our new notation
\begin{eqnarray}\label{eqFNd}
&&F_{dd} = 2Z_{dN^*}\tau_{N^*}F_{N^*d}\,, \\
&&F_{N^*d} = Z_{N^*d}\tau_dF_{dd}+Z_{N^*N^*}\tau_{N^*}F_{N^*d}\,,
\label{eqFNda} \\
&&F_{dN^*} = 2Z_{dN^*}\tau_{N^*}F_{N^*N^*}\,, \label{c} \\
&&F_{N^*N^*} = Z_{N^*d}\tau_dF_{dN^*}+Z_{N^*N^*}\tau_{N^*}F_{N^*N^*}\,.
\label{d}
\end{eqnarray}
A close examination of (\ref{eqFNd}) through (\ref{d}) reveals that we have two
decoupled sets of coupled equations sharing the identical integration kernel.
Indeed, the coupled equations (\ref{eqFNd}) and (\ref{eqFNda}) are transformed
into (\ref{c}) and (\ref{d}) by the substitutions
$F_{dd}\rightarrow F_{dN^*}$ and $F_{N^*d}\rightarrow F_{N^*N^*}$.
Therefore, it is sufficient to consider only one set for which we choose
the second one. After inserting (\ref{c}) into (\ref{d}) the former equation
may be written in the closed form
\begin{equation}\label{eqFNN}
F_{N^*N^*}=
(Z_{N^*N^*}+2Z_{N^*d}\tau_dZ_{dN^*})\,\tau_{N^*}F_{N^*N^*}. \\
\end{equation}
It is diagrammatically presented in Fig.~\ref{fig1}.
In analogy to a homogeneous Lippmann-Schwinger equation, we see that the
driving term $Z_{N^*N^*}$ plays
the role of a meson exchange $NN^*$-potential while the second term in
brackets gives the mechanism associated with intermediate
$NN$-interaction where the $\eta$ acts as a spectator.

\section{Two-body ingredients}\label{sect2}

Now we will specify the separable $\eta N$ and $NN$ scattering matrices
which determine the driving two-body forces in our model.
Since we restrict the pair-interaction to $s$-waves only,
the vertex functions
have a simple structure determined by two parameters,
the coupling constant $g_k$ and the cut-off $\beta_k$
\begin{equation}\label{eqFF}
\langle \vec p\,|f_k\rangle=f_k(p)=g_k\,F_k(p)
\,\mbox{ with }\, F_k(p)=\frac{\beta_k^2}{\beta_k^2+p^2}\,,
\end{equation}
where $\vec{p}$ denotes the relative momentum of the interacting pair.

For the $s$-wave $t$-matrix of the $NN=d$ isobar
\begin{equation}\label{eqTNN}
t_d(p,p',E_d)=f_d(p) \tau_d(E_d) f_d(p')\,,
\end{equation}
the following parametrization has been used
\begin{equation}\label{eqGNN}
g_d^2=\frac{16\pi a}{a\beta_d-2}\,, \quad
\gamma_d=-\frac{1}{2M_N}\,,
\end{equation}
where $a$ is the $NN$-scattering length.
For the propagator $\tau_d$ one then obtains
\begin{equation}\label{eqTaud}
\tau_d(E_d)=-\frac{1}{2M_N}\,
\Big[1+\frac{g_d^2\beta_d^3}
{16\pi\left(i\beta_d+\sqrt{M_N(E_d-2M_N)}\right)^2}
\Big]^{-1}\,.
\end{equation}
The $NN$-interaction parameters were taken from the
low-energy $np$-scattering fit of Yamaguchi
\cite{Yam}
\begin{equation}
\beta_d=1.4488\,\mbox{fm}^{-1},
\quad a=\left\{\begin{array}{rl}
5.378\,\mbox{fm}& \mbox{for the $^3S_1$-state,}\cr
-23.690\,\mbox{fm}& \mbox{for the $^1S_0$-state.}\cr
\end{array}\right.
\end{equation}

Analogously, we use for the $\eta N$=$N^*$ $t$-matrix the following
ansatz
\begin{equation}\label{eqTS11}
t_{N^*}(p,p',E_{N^*})=
f^{(\eta)}_{N^*}(p)\tau_{N^*}(E_{N^*})f^{(\eta)}_{N^*}(p') \,.
\end{equation}
Here the $N^*$-propagator is given in the form
\begin{equation}\label{eqTauN}
\tau_{N^*}(E_{N^*})=\Big(E_{N^*}-M_0-\Sigma_\pi(E_{N^*})-
\Sigma_\eta(E_{N^*})\Big)^{-1}\,,
\end{equation}
where $M_0$ denotes the bare mass of the $S_{11}$(1535)-resonance.
The resonance self energies associated with the couplings to the
$\pi N$ and $\eta N$ channels are determined by
\begin{equation}\label{eqSE}
\Sigma_j(E_{N^*})=\frac{1}{2\pi^2}\int\limits_0^\infty
\frac{f^{(j)2}_{N^*}(p)}{E_{N^*}-E_N(p)-\omega_j(p)+i\epsilon}\
\frac{p^2\,dp}{2\,\omega_j(p)}\,,
\end{equation}
where $\omega_j(p)=\sqrt{m_j^2+p^2}$ denotes the energy of
meson ``$j$'' and $E_N(p)=M_N+p^2/2M_N$ is the nonrelativistic total
nucleon energy. The meson-$N^*$ vertices are defined by $(j=\pi,\eta)$
\begin{equation}\label{eqFi}
f^{(j)}_{N^*}(p)=g^{(j)}_{N^*}\,
F_{N^*}^{(j)}(p)\,, \quad \mbox{with}\quad
F_{N^*}^{(j)}(p)=\frac{\beta_{N^*}^{(j)2}}{\beta_{N^*}^{(j)2}+p^2}\,.
\end{equation}
The $N^*$ parameters were chosen in such a way that the main ratios of the
hadronic decays of the $S_{11}$(1535) resonance are reproduced. We have
taken
\begin{equation}\label{eqBeTa}
g_{N^*}^{(\eta)}=2.0\,,\ g_{N^*}^{(\pi)}=1.5\,,\
\beta_{N^*}^{(\eta)}=6.5\,\mbox{fm}^{-1},\
\beta_{N^*}^{(\pi)}=4.5\,\mbox{fm}^{-1}.
\end{equation}
The bare $S_{11}$ mass $M_0$ was determined by the condition
\begin{equation}\label{eqM0}
M_0+\Re e\Big(\Sigma_\pi(M^*)+\Sigma_\eta(M^*)\Big)=M^* \,,
\end{equation}
where $M^*=1535$ MeV is the mass of the dressed isobar. The choice
(\ref{eqBeTa}) gives for the total and partial decay widths at the
resonance position
\begin{equation}
\Gamma=150\,\mbox{MeV}, \quad \frac{\Gamma_{\eta N}}{\Gamma}=
\frac{\Gamma_{\pi N}}{\Gamma}=0.5\,,
\end{equation}
which is reasonably consistent with the values given in the 1998 Particle
Data Group listings \cite{PDG98}.

For the actual evaluation of (\ref{eqFNN})
we use a natural representation in which
the total angular momentum $J$ and the total isospin $T$ are
diagonal.  The $\eta NN$ wave function is expanded into a
complete set of the following basis states
\begin{eqnarray}\label{eqJT}
|\vec{q}\,;\,JM_J\,TM_T\rangle&=& |\{(\tau_i\tau_j)t_k\tau_k\}TM_T\rangle\,
\sum\limits_{LM_L}\sum\limits_{SM_S}C_{LM_L\,SM_S}^{JM_J}\,
|\{(\sigma_i\sigma_j)s_k\sigma_k\}SM_S\rangle\,
|q;\,LM_L\rangle\, Y_{LM_L}(\hat{q})\,.
\end{eqnarray}
Here $\sigma_i$ and $\tau_i$ denote the spin and isospin of the individual
particles forming the isobar with the corresponding quantum
numbers $s_i$ and $t_i$. The partial waves are normalized as
\begin{equation}\label{eqNorm} \langle
q;\,LM_L|q';\,L'M'_L\rangle=
\frac{1}{q^2}\delta(q-q')\delta_{L'L}\delta_{M'_LM_L}\,.
\end{equation}
In (\ref{eqJT}) we have
already taken into account that the two-particle forces act only in
$l=0$ states. The momentum
$\vec{q}$ characterizes the relative motion of the correlated
$(ij)$-pair and a spectator particle. It is projected onto the partial
waves specified by the total orbital momentum $L$ with
projection $M_L$. Taking into account the parametrization adopted for
the two-body vertices, the expansion (\ref{eqJT}) turns the
operator equation (\ref{eqFNN}) into a set of one-dimensional
integral equations for the partial waves $|q;\,LM_L\rangle$.
In the following we consider only states with total orbital angular
momentum $L=0$. Then the two allowed configurations are
$(J^\pi;T)=(0^-;1)$ and $(1^-;0)$. Consequently, only the $^1S_0$
$NN$-state contributes to the three-body state $(0^-;1)$, and
correspondingly the $^3S_1$ $NN$-state to $(1^-;0)$.

In order to apply the representation (\ref{eqJT}) to Eq.~(\ref{eqFNN}),
we only need the spin-isospin recoupling coefficients
between the states $j$ and $k$ for which one has
\begin{eqnarray}\label{eqRcpl}
\langle\{(\sigma_i\sigma_j)s_k\sigma_k\}SM_S
|\{(\sigma_k\sigma_i)s_j\sigma_j\}S'M'_S\rangle &=&
\delta_{S'S}\delta_{M'_SM_S}\sqrt{(2s_j+1)(2s_i+1)}
W(\sigma_i \sigma_k \sigma_j S;\,s_js_k)\,, \\
\langle\{(\tau_i\tau_j)t_k\tau_k\}TM_T
|\{(\tau_k\tau_i)t_j\tau_j\}T'M'_T\rangle &=&
\delta_{T'T}\delta_{M'_TM_T}\sqrt{(2t_j+1)(2t_i+1)}
W(\tau_i \tau_k \tau_j T;\,t_jt_k)\,,
\end{eqnarray}
where $W(j_1\,j_2\,j_3\,j_4\,;\,j_5\,j_6)$ denotes the standard Racah
coefficient.

Taking the actual quantum numbers of the participating particles,
we obtain the following explicit form of the driving
terms appearing in (\ref{eqFNN})
\begin{equation}\label{eqZNN}
\langle
\vec{p};\,JT|Z_{N^*N^*}|\vec{p}\,';\,JT\rangle= V_\eta(p,p',W)+\chi
V_\pi(p,p',W)\,.
\end{equation}
Here the spin-isospin coefficients are $\chi=-1/3$ for
$(0^-;1)$ and $\chi=1$ for $(1^-;0)$. The meson-exchange potential
acting in the $L=0$ wave reads
\begin{equation}\label{eqVi}
V_j(p,p',W)= \frac{g_{N^*}^{(j)2}}{8\pi}
\int\limits_{-1}^{+1}\frac{
F_{N^*}^{(j)}\Big(\vec{p}\,'+\frac{M_N}{M_N+m_j}\vec{p}\Big)
F_{N^*}^{(j)}\Big(\vec{p}+\frac{M_N}{M_N+m_j}\vec{p}\,'\Big)}
{2\omega_j(|\vec{p}+\vec{p}\,'|)\,
\Big(W-E_N(p)-E_N(p')-\omega_j(|\vec{p}+\vec{p}\,'|)\Big)}\,
d(\hat{p}\cdot\hat{p}')\,,
\end{equation}
with $(j=\pi,\eta)$.
For simplicity we use the nonrelativistic relative meson-nucleon momenta
in the arguments of the regularization vertex form factors.
The driving term associated with the intermediate $NN$-interaction has
the form, using $E_d=W-\omega_{\eta}(q)-q^2/4M_N$ as the total c.m.\ energy
of the interacting $NN$-pair,
\begin{eqnarray}\label{eqZd}
\langle \vec{p}\,;\,JT|Z_{N^*d}\tau_d Z_{dN^*}|\vec{p}\,';\,JT\rangle
&=&
V_d(p,p',W) \nonumber \\&=&
\frac{2}{\pi}\int\limits_0^\infty
\frac{q^2\,dq}{2\omega_\eta(q)}\,V_N(p,q,W)\,
\tau_d\Big(W-\omega_{\eta}(q)-q^2/4M_N\Big)\,V_N(p',q,W)\,,
\end{eqnarray}
where the nucleon-exchange potential is given by
\begin{equation}\label{eqVN}
V_N(p,p',W)= \frac{g_{N^*}^{(\eta)}g_{d}}{8\pi} \int\limits_{-1}^{+1}
\frac{F_{N^*}^{(\eta)}\Big(\vec{p}\,'+\frac{m_\eta}{M_N+m_\eta}\vec{p}\Big)
\,F_d\Big(\vec{p}+\frac{\vec{p}\,'}{2}\Big)}
{W-E_N(p)-\omega_\eta(p')-E_N(|\vec{p}+\vec{p}\,'|)}\,
d(\hat{p}\cdot\hat{p}')\,.
\end{equation}

The integrals in (\ref{eqVi}) and (\ref{eqVN}) can be evaluated analytically.
The corresponding expressions, being rather cumbersome, are listed
in the appendix.
Finally, we present the partial wave representation of our basic
homogeneous equation (\ref{eqFNN})
\begin{equation}\label{eqFL}
F_{N^*N^*}(p,W)=\frac{2}{\pi}\int\limits_0^\infty
V_{N^*N^*}(p,p',W)\,
\tau_{N^*}(W-E_N(p')-p^{\prime\,2}/2M^*)\,
F_{N^*N^*}(p',W)\,p'\,^2dp'\,,
\end{equation}
where the notation
\begin{equation}\label{eqVL}
V_{N^*N^*}(p,p',W)=
2V_d(p,p',W)+V_\eta(p,p',W)+\chi\, V_\pi(p,p',W)
\end{equation}
is introduced, and the
invariant mass of the $N^*$-isobar in (\ref{eqVL}) was evaluated as
$E_{N^*}=W-E_N(p')-p'^2/2M^*$.
Anticipating a later result, we note that the contribution from
the pion exchange potential $V_\pi$ in (\ref{eqVL})
is practically insignificant, and almost the
whole attraction in the $\eta NN$-system comes from the first two terms with
approximately equal strengths.

The method for searching the zeros of the Fredholm
determinant $D(W)$ is to approximate the integral in
Eq.~(\ref{eqFL}) by a finite sum, transforming it into an ordinary
matrix equation. Then (\ref{eqD}) may be written as an
algebraic equation
\begin{equation}\label{eqDL}
\det\Big|\delta_{ij}-\frac{2}{\pi}p_j^2C_j
V_{N^*N^*}(p_i,p_j,W)
\tau_{N^*}\Big(W-E_N(p_j)-\frac{p_j^2}{2M^*}\Big)
\Big|=0\,.
\end{equation}
Here $C_j$ are the weights for the chosen quadrature (in the present
calculation we have chosen the Gauss quadrature).

\section{Structure of the Riemann surface and continuation into
nonphysical sheets}\label{sect3}

\begin{figure}
\centerline{\psfig{figure=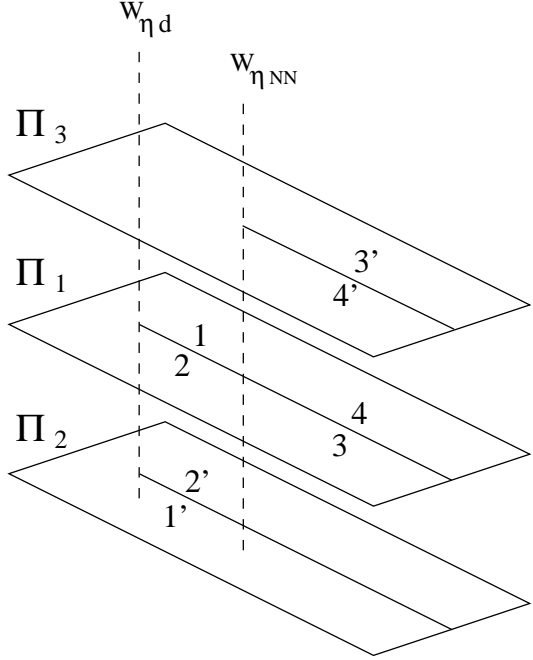,width=7cm,angle=0}}
\vspace{.5cm}

\caption{
Structure of the Riemann surface for the $\eta NN$ scattering matrix
in the $(1^-;0)$ channel in the near threshold region.
Shown are the physical sheet
$\Pi_1$ and the adjacent nonphysical sheets $\Pi_2$, and $\Pi_3$.
Two vertical dashed lines pass through the two-body ($W_{\eta d}$) and
three-body ($W_{\eta NN}$) thresholds. The numbers $i$ and $i^\prime$
indicate the
identified rims of the associated two- and three-body unitary cuts.
}
\label{fig2}
\end{figure}

Our main concern is to find the zeros of the Fredholm determinant
in the complex energy plane resulting in poles of the scattering
matrix. The structure of the manifold Riemann surface for the $\eta
NN$-system in $(J^\pi;T)=(1^-;0)$ channel is presented in Fig.~\ref{fig2}.
Shown are the physical sheet $\Pi_1$ and the nearest nonphysical
sheets $\Pi_2$, $\Pi_3$ directly adjacent to $\Pi_1$.
The sheet $\Pi_1$ has a conventional analytical structure.
Namely, it may have poles corresponding to a possible formation of
three-body bound states, and the unitarity right-hand cuts.
The former stem from the Cauchy type integrals inherent in the
integration kernel in (\ref{eqFL}).
There are two main cuts which determine the
structure of the Riemann surface for the three-body problem
in the energy region of interest:

(i) A two-particle cut, beginning at the two-body threshold
$W_{\eta d}=2M_N+m_\eta+\epsilon_d$ where $\epsilon_d<0$ is the
deuteron binding energy.
The cut arises from the propagator $\tau_d(E_d)$
which has a pole when $E_d-2M_N=\epsilon_d$.
This pole is associated with the elastic $\eta$-deuteron scattering
and gives the well known two-body
contribution to the common three-body unitarity relations. The
corresponding square-root branch point $W=W_{\eta d}$ develops the
two-sheet structure $\{\Pi_1,\,\Pi_2\}$ typical for conventional
two-body scattering.

(ii) A three-body cut, starting at the three-body threshold $W_{\eta
NN}=2M_N+m_\eta$. This cut is induced by the two-body right-hand cut
in $\tau_d$ corresponding to the $d\to NN$ break-up
and by the similar cut in $\tau_{N^*}$ associated with the decay
$N^*\to \eta N$. The corresponding nonphysical sheet is denoted as $\Pi_3$
in Fig.~\ref{fig2}.
Furthermore, the driving terms $V_\pi$, $V_\eta$,
and $V_N$ in (\ref{eqVi}) through (\ref{eqVN}) contain the well known
logarithmic
singularities which are analogous to the dynamical (left-hand)
singularities of the nucleon-nucleon OBE potential and correspond to the
exchange of a real particle which becomes possible when $W\geq W_{\eta
NN}$. For some values of $p$ and $p'$ these singularities pinch the real
axis producing the additional three-body cut beginning at the branch point
$W=W_{\eta NN}$. In the partial wave representation, this point
is of logarithmic type and gives rise
to an infinite number of sheets at $W=W_{\eta NN}$. The structure
of the Riemann surface associated with the logarithmic singularities
is in itself of no physical interest and is therefore not presented
in Fig.~\ref{fig2}.

All sheets are connected as depicted in Fig.~\ref{fig2}. The sheet $\Pi_2$
may contain the poles which show-up as correlated two-body $\eta d$
states (virtual or resonant). The analogous states which may
be observed in the three particle scattering processes
are located on the sheet $\Pi_3$.
In the absence of an actual three-body scattering
experiment, these states may occur as final states in
reactions with, e.g., deuteron break-up, such as $\gamma d\to\eta
np$.  Following the terminology accepted in the literature we call
$\Pi_2$ and $\Pi_3$ as two-body and three-body sheets, respectively.

The singularity structure outlined above arises from the free three-body
propagators and the poles associated with the bound states of
two-body subsystems. It is also presented in detail in
Refs.~\cite{Orlov,Glock,Moll,Kolg,Bel} in relation to the
three-nucleon problem.
The Riemann surface for the $\eta NN$ scattering matrix is
formally more complicated due to the following reasons:

(i) Since there is a rather strong coupling between the
$\eta N$ and $\pi N$ channels in the energy region of the $S_{11}$(1535)
resonance, all sheets depicted in Fig.~\ref{fig2} have an additional cut
beginning at $\pi NN$ threshold\footnote{\small
Due to the spin-isospin selection
rules the $\pi d$ channel does not appear in the configurations
$(J^\pi;\,T)=(1^-;\,0)$ and $(0^-;\,1)$ considered in this paper.
Therefore no additional two-body cut starting at the $\pi d$ threshold
is present.}.
The typical structure
of the Riemann surface associated with the $\eta N\leftrightarrow\pi N$
coupling is presented in Fig.~\ref{fig3}.

(ii) The three-body sheet $\Pi_3$ has an additional
square-root branch point associated with the
quasi-two-body $NN^*$-threshold. Treating the bare $N^*$ mass as noted in
Sect.~\ref{sect3} (see Eq.~(\ref{eqM0})), we keep the corresponding complex
cut well away from the relevant energy region and, therefore, will
ignore it in the following considerations.

The structure of the Riemann sheet for the $\eta NN$ scattering
matrix in the $(J^\pi;T)=(0^-;1)$ state is different from that for
$(J^\pi;\,T)=(1^-;\,0)$. Namely, due to the virtual
character of the $^1S_0$ pole, the corresponding two-body sheet
$\Pi_2$ is "glued" not directly to $\Pi_1$ as previously
but to the three-body sheet $\Pi_3$.

From ordinary potential scattering theory it is known that the poles of
the scattering matrix corresponding to virtual or resonant
states are located in the nonphysical energy domain.
The continuation into
this area may be performed by going around the branch point at threshold
in order to get the virtual pole or directly from the upper half
complex energy plane by crossing the real positive axis to reach the
resonance pole. This continuation may clearly be done if the
analytical expression for the scattering matrix is available. Otherwise,
one has to use the dynamical equation written in terms of the real energies
in order to continue it into the nonphysical area. In the former
case, several recipes may be used (for a review of the methods see, e.g.
\cite{Orlov}).

Turning to the three-body problem we would like to
mention the method of analytical continuation based on a contour
deformation technique. This method,
developed in \cite{Glock,Moll}, was previously applied to
the three-neutron problem and later to the study of the $\Sigma^-nn$- and
$\Lambda nn$-interaction \cite{Mats}. In this paper we extend this
technique to the $\eta NN$-interaction
including the inelastic $\pi NN$-channel. Firstly, we would like to
review briefly the basic details of the method. Let us
consider the following Fredholm equation
\begin{equation}\label{eqFpz}
F(p,W)=\int\limits_0^\infty
\frac{f(p',W)}{2mW-p^2-p'^2-(\vec{p}+\vec{p}^{\,\prime})^2}F(p',W)dp'\,,
\end{equation}
where the function $f(p',W)$ does not have any singularity in the
relevant energy domain. For simplicity we consider three
particles having equal masses $m$ and use nonrelativistic
kinematics. The energy $W$ is the total kinetic energy. The physical
region is determined in the usual fashion: $\Re e\,W>0$,
$\Im m\,W=\varepsilon\to +0$.
Because of the singularities of the kernel for real
$W$ one can not directly continue the equation down through
the cut $0\leq W<\infty$.
In this case, shifting the integration path for $p'$ to the position
$C$ in Fig.~\ref{fig4}
(the angle $\theta$ is arbitrarily taken to be $\pi/4$)
we can cross the real energy axis and enter the area on the
nonphysical sheet. Here the variable $p$ is also taken
on the contour $C$. The parabolic borderline separates the analytical
domain from the forbidden area where the denominator in
(\ref{eqFpz}) may vanish for some values of $p$ and $p'$. If one uses
relativistic kinematics the permissible area slightly narrows. The
continuation procedure is of course justified when the contour being
deformed does not hit any singularities of the integration kernel. It
must be noted that the method allows one to uncover only the certain part
of the nonphysical sheet. However, varying suitably the contour parameters
$(a,\theta)$, one can get the major part of the nonphysical
domain. We note also that in the present calculation
the procedure described above was applied directly to
Eq.~(\ref{eqDL}), since all singularities of the kernel are
contained in the Fredholm determinant $D(W)$.

\subsection{Continuation into the lower half of the sheet
$\Pi_1$}\label{subsect51}

\begin{figure}
\centerline{\psfig{figure=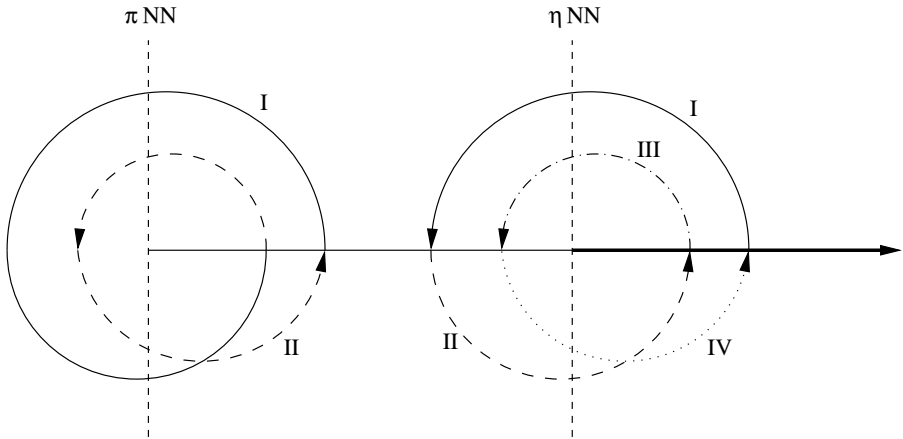,width=12cm,angle=0}}
\vspace{.5cm}
\caption{
Relationship of the four sheets for a two-channel $\eta NN$-$\pi NN$
problem. The point moving around the $\eta NN$ threshold passes
successively all four sheets denoted here by the roman numerals.
}
\label{fig3}
\end{figure}

\begin{figure}
\centerline{\psfig{figure=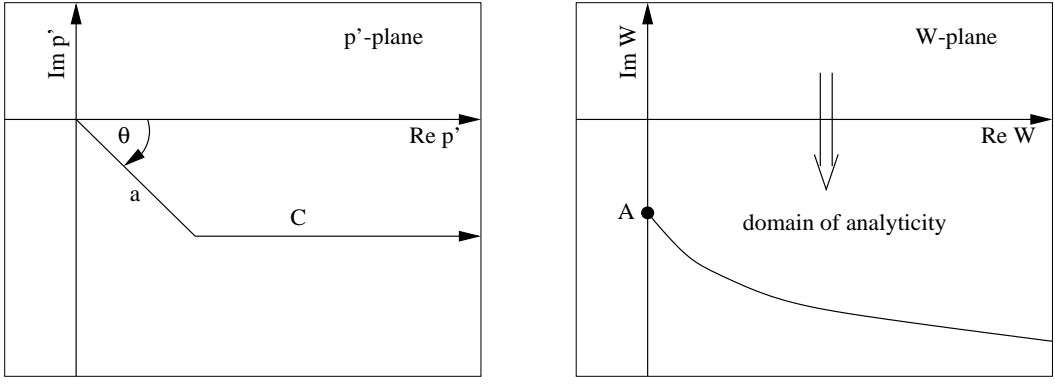,width=14cm,angle=0}}
\vspace{.5cm}
\caption{
Analytical continuation of the integral equation (\protect{\ref{eqFpz}})
within the contour deformation method. The left-hand panel shows
the new integration path $C$, determined by the length of a finite straight
line $a$ and a rotation angle $\theta$ (here $\theta=\pi/4$). On the
right-hand  panel, the new domain of analyticity in the lower half-plane
of the nonphysical sheet is shown. The energy, corresponding to the
corner point $A$, may be evaluated as $W_A=0-ia^2/m$.
}
\label{fig4}
\end{figure}

From the diagrams shown in Fig.~\ref{fig3} it
is clear that the lower half-plane
of the sheet $\Pi_1$ in Fig.~\ref{fig2}
is at once the lower half of the three-body
nonphysical sheet for the $\pi NN$ channel reached from above by
crossing the cut between $\pi NN$ and $\eta d$ thresholds. This area
is indicated as (II) in Fig.~\ref{fig3}. Therefore, when
continuing into this domain, the logarithmic singularities of the three-body
propagator in $V_\pi(p,p',W)$ present an obstacle. A slight shift of
the integration path (analogously to the pattern in Fig.~\ref{fig4})
into the fourth quadrant allows one to reach the area shown
in Fig.~\ref{fig5}.
Some calculational problems within this method may occur when the energy
$W$ approaches the real axis above the $\eta NN$ three-particle threshold
since the analyticity domain degenerates into part of a straight line.
In this limit one has to handle the numerics with greater accuracy.

\subsection{Continuation into the sheet $\Pi_2$}\label{subsect52}

In order to continue Eq.~(\ref{eqDL}) into the sheet $\Pi_2$, we adopt
the following procedure. Entering into this sheet is
accompanied by the movement of the pole $p_0$ of the propagator
$\tau_d$ into the fourth quadrant of $p'$. This requires the
contour deformation as is shown on Fig.~\ref{fig6}.
In restoring the integration
path to its original position, we pick up the new term in
the potential $V_d(p,p',W)$ corresponding to the residue of the
integrand in (\ref{eqZd}) at $q=p_0$
\begin{eqnarray}
V_d(p,p',W)&=&\frac{2}{\pi}\int\limits_0^\infty\frac{q^2\,dq}
{2\omega_\eta(q)}
V_N(p,q,W)\,\tau_d\Big(W-\omega_{\eta}(q)-q^2/4M_N\Big)\,V_N(p',q,W)
\nonumber \\
&&+i \,\frac{g_{d}\,\beta_d^{3/2}}{2\,\sqrt{\pi}}\,
\frac{p_0\sqrt{\frac{|\epsilon_d|}{M_N}}}{\omega_\eta(p_0)+2M_N}\,
V_N(p,p_0,W)\,V_N(p',p_0,W)\,,
\end{eqnarray}
where
\begin{equation}
p_0=2\,\sqrt{M_N}\,\Big\{W+|\epsilon_d|-\sqrt{
m_\eta^2+4M_N(W+|\epsilon_d|-M_N)}\Big\}^{1/2}\,,
\end{equation}
(compare with (\ref{eqZd})). In this manner we may cross the
two-body cut in the interval $W_{\eta d} < W < W_{\eta NN}$
and enter into the lower half-plane of the sheet $\Pi_2$
(transition $2\to 2'$ in the notation of Fig.~\ref{fig2}).
The new domain of analyticity is shown in Fig.~\ref{fig6}.

Continuation into the upper half-plane of the sheet $\Pi_2$
(transition $1\to 1'$) may be performed in a like manner. In this case,
the pole $p_0$ crosses the integration contour when moving from below
into the first quadrant of the $p'$-plane.

\subsection{Continuation into the sheet $\Pi_3$}\label{subsect53}

For the continuation into the sheet $\Pi_3$ from the region above the
segment $W_{\eta NN}\leq W<\infty$ of the real energy axis, we use
the same technique as was outlined in the case A. In order
to reach the domain below the $\eta NN$ threshold it is necessary to take
$\theta > \pi/4$.  In this case, however, one has to avoid the singularities
on the imaginary $p'$-axis which come from the regularization
form factors as well as from the factors of the type
$1/\sqrt{2\omega_j(p)}$ in the integrands in (\ref{eqVi}).
Therefore, the rotation angle $\theta$ must be restricted to values
less than $\pi/2$. Thereby, we also do not encounter any problem with the
poles of $\tau_d$ and $\tau_{N^*}$.  Choosing the mesh points from the
deformed contour $C$,
the three-body energy $W$ can be taken to pass above the
three-body threshold $W_{\eta NN}$ and down through the cut into the
lower half-plane of the sheet $\Pi_3$ as is shown in Fig.~\ref{fig7}
(transition $4\to 4'$ in Fig.~\ref{fig2}).

The analytical continuation
into the upper half-plane of the sheet $\Pi_3$ is easily possible if the
pion-exchange potential is ignored. Otherwise the logarithmic singularities
from $V_\pi$ bar the continuation path. For this reason, when entering
this area from the lower half-plane of $\Pi_1$, we will switch off
the pion-exchange forces. In this case, in order to make the transition
$3\to 3'$ in the notation of Fig.~\ref{fig2}, one just has to shift the
contour into the first quadrant of the integration variable.
Taking into account the smallness of the contribution from $V_\pi$ as
denoted above, we believe that neglection of the pion-exchange potential
does not spoil significantly the quality of our results.

\section{Results and discussion}\label{sect6}

In order to obtain a better insight into the $\eta NN$ low-energy
dynamics, we have investigated the trajectories drawn by the $S$-matrix poles
on the Riemann surface when varying the interaction parameters. In
doing this, we are guided by the following consideration.  We presume
that the pole trajectory, on which one expects to find a resonance or
virtual three-body state closest to threshold, will also contain the
deepest bound state eigenvalue. In this regard, we artificially enhanced
the strength of the $\eta NN$-forces until the first bound state (i.e., the
pole on the sheet $\Pi_1$) appears. Then we have approached the
actual physical situation by weakening the interaction strength to
the value determined by the parameters given in Sect.~\ref{sect2}.
Following this variation, the pole moves on the Riemann surface and
finally arrives at
several positions, developing in this way the bound (virtual) or
resonant three-body state. Clearly, it may also appear on a Riemann
sheet far removed from the physical region and thus will not influence
the real scattering processes.

Turning back to the $\eta NN$-system, we take as a varying
parameter the coupling constant $g_{N^*}^{(\eta)}$. As its
physical value we consider that presented in (\ref{eqBeTa}). We will
start with the $(1^-;0)$ channel, the quasideuteron case.
Before proceeding further, in order to see how the pole
may behave when
approaching the threshold region, we will ignore for the moment being the
$\pi NN$-channel by setting $g_{N^*}^{(\pi)}$=0.
Then we took $g_{N^*}^{(\eta)}$=4  and
found the bound state pole at $W_{pol}=W_{\eta NN}-8.08$ MeV.  As
expected, the weakening of $g^{(\eta)}_{N^*}$ implies the motion of
the pole towards the threshold region.  At $g^{(\eta)}_{N^*}$=2.5  it
overtakes the $\eta d$ two-body threshold and passes into the
two-body sheet $\Pi_2$ producing a virtual $\eta d$ state. Further
weakening of the $\eta N$-attraction pulls the pole back along the
negative real axis away from the physical region. We would like to note
that this shape of the trajectory is what we can expect
naively from ordinary two-body potential scattering. Owing to the
absence of the centrifugal barrier, the $s$-wave pole develops a virtual
state and not a resonance when passing into the nonphysical sheet
(here we do not touch upon exotic cases like the multitude of
$s$-wave resonances in a deep square well \cite{Tay}). We may expect to
encounter the same pole behaviour in a three-body case since our physical
situation is in effect the two-particle interaction with one being complex.

Now including again the $\pi NN$ channel, we see that its effect is to
shift the starting position of the pole downwards from the real axis
and thus to transform the real $\eta NN$ bound state into the $\pi
NN$ resonance (or, what amounts to the same, the $\eta NN$ bound state
with a finite lifetime). Then decreasing
$g_{N^*}^{(\eta)}$ and allowing the eigenvalue $W_{pol}$ to pass
the $\eta NN$ threshold, we reach the situation illustrated in Fig.~\ref{fig8},
where some pole trajectories corresponding to the different values of
$g^{(\pi)}_{N^*}$ are shown. One can see that the widths of the
hypothetical $\eta NN$ bound states are sufficiently less than that
of the $S_{11}$(1535) resonance in this energy region. This is due
partially to the fact that $\eta NN$-system spends a large fraction of
time in the $\eta (NN)$ state (not in $N(N^*)$), which in this region
has no direct coupling with any configuration in the continuum.
This effect explains also the anomalous decrease of the $\pi NN$
resonance width when approaching the $\eta$-production threshold.
As one can see from Fig.~\ref{fig8}, with
increasing $g^{(\pi)}_{N^*}$ the trajectories are shifted
downwards and the point where the pole meets the real
energy axis moves to the right. For the physical value
$g^{(\pi)}_{N^*}$=1.5 this point is located
at $W \approx W_{\eta NN}$+1.51 MeV.
The trajectory crosses the real axis
above the three-body threshold and passes to the
three-body sheet $\Pi_3$ (dashed curve with open triangles
in Fig.~\ref{fig8}).
If one go around the three-body threshold
$W_{\eta NN}$ and enters the upper
half plane of the two-body sheet $\Pi_2$ (dotted line in
Fig.~\ref{fig8}), one finds the extension of the
trajectory also into this sheet (solid curve with filled circles).
We recall that we neglected the pion-exchange potential when calculating
the trajectory passing into the sheet $\Pi_3$
(see Sect.~\ref{subsect53}). For this reason the
parts of the last two trajectories lying on the sheet
$\Pi_1$ are not identical.
We see however that the difference is not significant.
The pole positions corresponding to
$g^{(\eta)}_{N^*}$=2.0 are $W_{pol}=W_{\eta NN}-(3.00-i13.67)$ MeV on the
sheet $\Pi_3$ and $W_{pol}=W_{\eta d}-(4.39-i7.22)$ MeV on the sheet
$\Pi_2$.

After these findings we have carried out a careful
search for poles on the lower half-plane of the sheets
$\Pi_2$ and $\Pi_3$. Of course, we
were interested primarily in the energy area just beyond the real
axis where possible resonant states may occur.
Our conclusion is that no
poles appear in the physically interesting domain at least for
reasonable values of the interaction parameters.  As for the presence of a
possible
resonance pole on the three-body sheet $\Pi_3$, one can assume that
it may be found beyond the complex unitarity cut generated by
the $NN^*$ break-up singularity in $\tau_{N^*}$. In any case, this
pole, if it exists, is far removed from the relevant energy
region and will hardly influence the near threshold $\eta
NN$ interaction.
Thus within our model, the $S$-matrix poles nearest to the
physical region are situated on the upper half plane of
the sheets $\Pi_2$ and $\Pi_3$ and produce
virtual rather than resonant states in the $\eta NN$-system.

Now we will consider the $(J^\pi;T)=(0^-;1)$ channel. We make the same
variation of the $N^*$ coupling constants and obtain the trajectories
depicted in Fig.~\ref{fig9}.
Switching off the $\pi NN$ channel, we found
the corresponding trajectory moving around the $\eta NN$ threshold
and ending in the third quadrant of the variable $W-W_{\eta NN}$
on the sheet $\Pi_3$. The similar pole behaviour was noted also for
the $\Sigma^-nn$ and $\Lambda nn$ $s$-wave configurations in Ref.~\cite{Mats}.
Increasing $g_{N^*}^{(\pi)}$
and varying $g_{N^*}^{(\eta)}$ from 4 to 2, we
find the general picture to be qualitatively the same as in the
quasideuteron case. One notes a visible shift of
the trajectories towards the higher energies, which is most likely
explained by the weakening of the $NN$-interaction in the
singlet $^1S_0$ state in relation to the more attractive $^3S_1$ state.
The trajectory corresponding to
$g_{N^*}^{(\pi)}$=1.5 brings the final position of the pole to
the point $W_{pol}=W_{\eta NN}+(14.06+i17.19)$ MeV.

What are the physical consequences that we can extract from the results
above? First, as we just have shown, it is unlikely that a
low-energy resonance will be found in the $(0^-;1)$ or
$(1^-;0)$ channels. On the other hand, the virtual poles
on the sheets $\Pi_2$ and $\Pi_3$ may influence the value of the
scattering amplitudes through its proximity to the threshold region.
Thus the next logical step in this direction would be to
calculate, e.g., the $\eta d$-scattering near threshold. This
problem, when treated within a three-body approach,
is rather complex in itself and is beyond the scope of the
present paper. The corresponding investigations for the elastic and
inelastic $\eta d$-scattering will be presented elsewhere (see also
\cite{Shev,Garc}).
Here we adopt the approximate formula
for the $s$-wave elastic two-body scattering, when the amplitude has a virtual
pole near zero energy (see e.g.\ \cite{Land})
\begin{equation}\label{sigED}
\sigma_{\eta d}=\frac{4\pi}{|p_{pol}-p|^2}\,,
\end{equation}
where $p$ denotes the c.m.\ $\eta d$ momentum and $p_{pol}$ its pole
position, determined as
\begin{equation}
\displaystyle
p_{pol}=\sqrt{\frac{2M_dm_\eta}{M_d+m_\eta}\,
(W_{pol}-M_d-m_\eta)}\,, \quad \Im m\,p_{pol}<0\,,
\end{equation}
with $M_d$ being the deuteron mass.
In Fig.~\ref{fig10} we show the result obtained for
$W_{pol}=W_{\eta d}-(4.39-i7.22)$ MeV.
One sees a strong
enhancement of the cross section as we approach the threshold energy, an
effect which is typical when the system
possesses a weak virtual state. Therefore we conclude that in the region
of a few MeV above threshold an anomalous
resonance-like behaviour can in general be expected in the
cross section of reactions with $\eta NN$-channels in the final
state.

Finally we must note that our results are in contrast to those
of Shevchenko et al.~\cite{Shev} who claimed the existence
of $s$-wave $\eta d$-resonances. They used the on-shell solution of
the nonrelativistic three-body equation and found the counterclockwise
rotation of the $\eta d$ scattering amplitude in the Argand plot.
The reason for this disagreement is not completely clear to us.
On the other hand, we would like to mention the qualitative agreement of our
results with those presented in \cite{Garc} where a similar behaviour
of the $\eta d$-scattering cross section in the low-energy region
was found.

\section{Summary and conclusion}

In order to answer the question whether there might exist a low-energy
resonance or virtual $\eta NN$-state, we have studied the typical pole
trajectories $W_{pol}(g_{N^*}^{(\eta)})$ for the $\eta NN$-system in
$L=0$ states.  Decreasing the $\eta N$-interaction parameter
$g_{N^*}^{(\eta)}$, the pole overtakes the three-body
threshold and moves into the upper half plane of the sheets $\Pi_2$ and
$\Pi_3$ adjacent to the lower rim of the two- and
three-body unitarity cuts on the physical sheet $\Pi_1$.
Since these areas are not directly
connected with the physical one, we have to conclude that no
three-body resonances can occupy these trajectories. Moreover, we
can assume that the very pattern of the pole trajectories forbids a
low-energy $\eta NN$ resonance independent of the actual values of
interaction parameters. We expect this feature to be valid in
general for ($L=0$)-$\eta NN$ configurations, at least within the model
of a type presented here.
On the other hand, our results point to a possible explanation of
the strong enhancement of the $\eta$-production cross section near
threshold. This effect may be assigned not to a resonance but to
virtual $\eta NN$-states, indeed generated by the poles on the sheets
$\Pi_2$ and $\Pi_3$.

We would like to emphasize that our conclusion is based essentially
on the model of a pure $s$-wave interaction. The typical feature of
an $s$-wave interaction known from the familiar two-body problem is to
develop a virtual rather than a resonance state.
In this context, it will be of interest to enrich the present study
by including $p$-waves in the partial wave decomposition (\ref{eqJT}).
It must, however, be kept in mind that $p$-waves are energetically far less
favourable in a low-energy regime. Therefore, we think that even though a
pole may come sufficiently close to threshold, quite a strong enhancement
factor from a $p$-wave contribution would be needed in order to make a
possible $p$-wave resonance visible in the cross section.

Another aspect, left out in the present work, is the so-called direct
$NN^*$-interaction which is not automatically accounted for in the
conventional three-body approach. Due to the short-range character
of this interaction, associated with its static nature, the corresponding
terms are expected to give only a small contribution when compared to the
full scattering amplitude (with respect to the role of these terms in
$\Delta N$-interaction see e.g.\ \cite{Arenh}). However, it should be
remembered, that three-body low-energy dynamics is known to be rather
sensitive to the model variation. Thus even a small perturbation may
change some of the present results.

\acknowledgments
One of the authors (A.\,F.) is grateful to
the theory group of the Institut f\"ur Kernphysik
at the Johannes Gutenberg-Universit\"at Mainz for the very kind hospitality.
Special thanks goes to Dr.\ Michael Schwamb for many fruitful
discussions.

\renewcommand{\theequation}{A\arabic{equation}}
\setcounter{equation}{0}
\section*{Appendix}
\label{app}

For the driving terms in (\ref{eqVi}) and (\ref{eqVN}) one finds the
following analytic expressions:

(a) meson exchange potential $(j=\pi,\eta)$:
\begin{eqnarray}
V_j(p,p',W)=&&
\frac{g_{N^*}^{(j)2}\beta_{N^*}^{(j)4}}{16\pi}
\Big(\frac{M_N+m_j}{M_N}\Big)^2
\frac{1}{p\,p'\,(\alpha_1^2+c^2)(\alpha_2^2+c^2)}\nonumber\\
&&\times \Big\{\ln{\frac{c-a_{-}}{c-a_{+}}}-
\Big[\frac{\alpha_2^2+c^2}{2\,(\alpha_1^2-\alpha_2^2)}\,
\Big(2\,\ln{\frac{|A_1^+|}{|A_1^-|}}
-i\,\frac{c}{\alpha_1}\,\ln{\frac{A_1^+\,A_1^-}{(A_1^+\,A_1^-)^*}}
\Big) + (1\leftrightarrow 2)\Big]\Big\}\,,
\end{eqnarray}
where
\begin{eqnarray}
c&=&W-E_N(p)-E_N(p')\,,\\
a_{\pm}^2&=&m_j^2+(p\pm p^{\prime\,})^2\,,\\
\alpha_1^2&=&\frac{M_N+m_j}{M_N}
\beta_{N^*}^{(j)2}-m_j^2+\frac{m_j}{M_N}p^{\prime\,2}-
\frac{m_j}{M_N+m_j}p^2\,,\\
\alpha_2^2&=&\frac{M_N+m_j}{M_N}
\beta_{N^*}^{(j)2}-m_j^2+\frac{m_j}{M_N}p^2-
\frac{m_j}{M_N+m_j}p^{\prime\,2}\,,
\end{eqnarray}
and $A_k^{\pm}=\alpha_k\pm i\,a_\pm$ for $k=1,2$.
If $\alpha_1=\alpha_2=\alpha$, then with $A^{\pm}=\alpha\pm i\,a_\pm$
\begin{eqnarray}
V_j(p,p',W)&=&
\frac{g_{N^*}^{(j)2}\beta_{N^*}^{(j)4}}{16\pi}
\left(\frac{M_N+m_j}{M_N}\right)^2
\frac{1}{pp'(\alpha_1^2+c^2)^2}\nonumber\\
&&\times \Big\{\ln\Big(
\frac{c-a_{-}}{c-a_{+}}
\,\frac{|A_1^+|}{|A_1^-|}\Big)+i\,c\,\frac{3\alpha^2+c^2}{4\alpha^3}
\,\ln{\frac{A^+\,A^-}{(A^+\,A^-)^*}}
+2\,\Re e \Big[\Big(i+\frac{c}{\alpha}\Big)
\frac{a_{-}-a_{+}}{A^+\,A^{-\,*}}\Big]\Big\}\,.
\end{eqnarray}

(b) nucleon exchange potential:
\begin{eqnarray}
V_N(p,p',W)&=&\frac{g_{N^*}^{(\eta)}g_d\beta_{N^*}^{(\eta)2}\beta_d^2}
{16\pi \,pp'}\frac{M_N(M_N+m_\eta)}{m_\eta}
\frac{1}{(a_2-a_1)(a_2+c)(a_1+c)}
\nonumber \\
&&\times
\left\{(a_2+c)\,\ln\frac{a_1+pp'}{a_1-pp'}-(a_1+c)\,\ln\frac{a_2+pp'}{a_2-pp'}-
(a_1-a_2)\,\ln\frac{c+pp'}{c-pp'}\right\}\,,
\end{eqnarray}
where
\begin{eqnarray}
a_1&=&\frac{M_N+m_\eta}{2m_\eta}
\left(\beta^{(\eta)2}_{N^*}+p^{\prime\,2}\right)+
\frac{m_\eta}{2(M_N+m_\eta)}p^2\,, \\
a_2&=&\beta_d^2+p^2+p^{\prime\,2}/4\,,\\
c&=&M_N(W-E_N(p)-\omega_{\eta}(p')-M_N)-(p^2+p^{\prime\,2})/2\,.
\end{eqnarray}

\newpage

\begin{figure}
\centerline{\psfig{figure=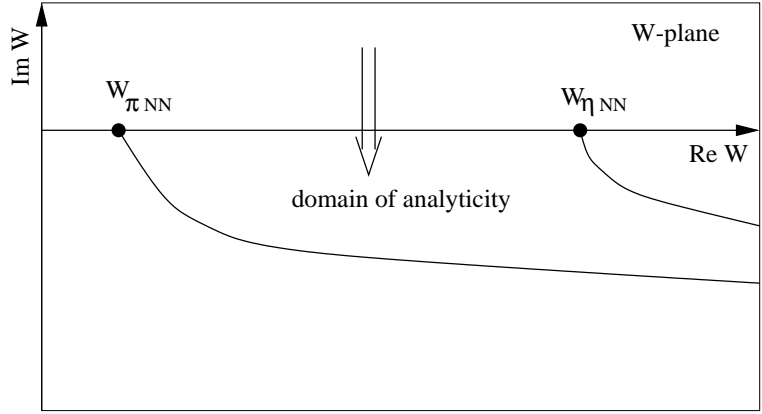,width=10cm,angle=0}}
\vspace{.5cm}
\caption{
The energy domain of the sheet $\Pi_1$ available for the analytical
continuation. The parameters of the integration path $C$
having the form shown in Fig.~\protect{\ref{fig4}} are:
$a$=200 MeV/c, $\theta=15^\circ$.
}
\label{fig5}
\end{figure}

\begin{figure}
\centerline{\psfig{figure=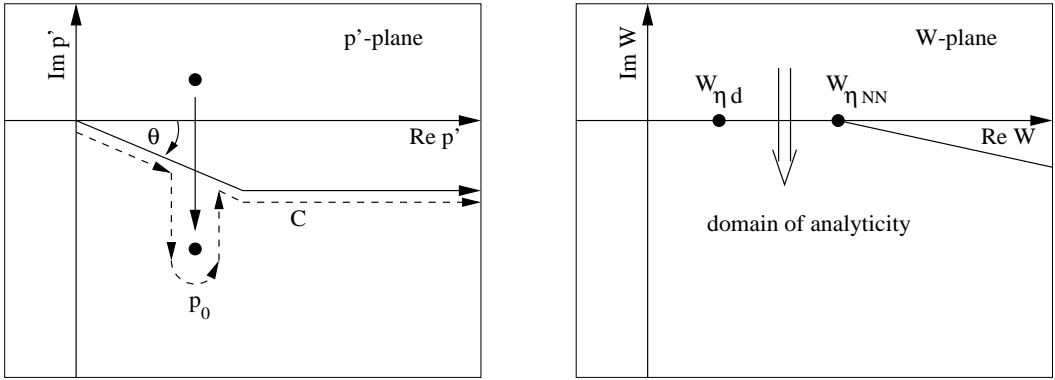,width=14cm,angle=0}}
\vspace{.5cm}
\caption{
Continuation into the two-body sheet $\Pi_2$. On the left-hand panel
we show the pole $p_0$
of the propagator $\tau_d$ moving across the
integration path when the energy $W$ passes through the two-body cut
on the interval $W_{\eta d} < W < W_{\eta NN}$. The parameters of the new
contour are the same as in Fig.~\protect{\ref{fig5}}. The energy domain
reached on the sheet $\Pi_2$ is shown on the right-hand panel.
}
\label{fig6}
\end{figure}

\begin{figure}
\centerline{\psfig{figure=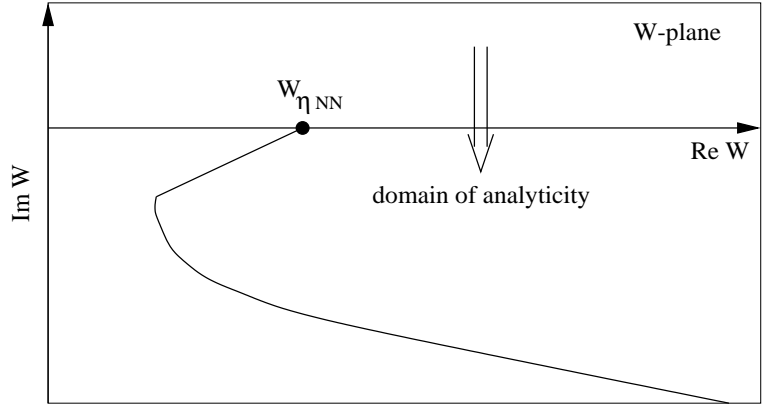,width=10cm,angle=0}}
\vspace{.5cm}
\caption{
The same as in Figs.~\protect{\ref{fig5}} and \protect{\ref{fig6}} for
the sheet $\Pi_3$. The parameters of the shifted contour are
$a$=200 MeV/c, $\theta=80^\circ$.
}
\label{fig7}
\end{figure}

\begin{figure}
\centerline{\psfig{figure=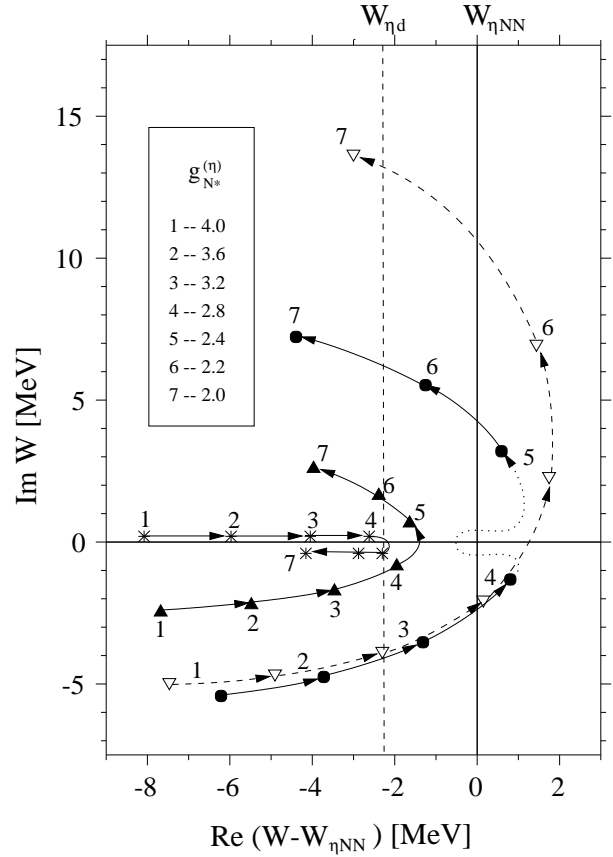,width=8cm,angle=0}}
\vspace{.5cm}
\caption{
The pole trajectories of the $\eta NN$ scattering matrix for the
$(J^\pi;T)=(1^-;0)$ state. Three trajectories correspond to the
different choices of the $\pi NN^*$ coupling constant:
$g_{N^*}^{(\pi)}$=0\,(stars), $g_{N^*}^{(\pi)}$=1\,(filled triangles),
and $g_{N^*}^{(\pi)}$=1.5\,(filled circles).
The legend shows the
values of the $\eta NN^*$ coupling constant $g_{N^*}^{(\eta)}$
being varied along the individual trajectory.
The trajectory shown by the dashed curve is continued into the three-body
sheet $\Pi_3$. The other trajectories pass into the two-body sheet
$\Pi_2$ (see the text).
}
\label{fig8}
\end{figure}

\begin{figure}
\centerline{\psfig{figure=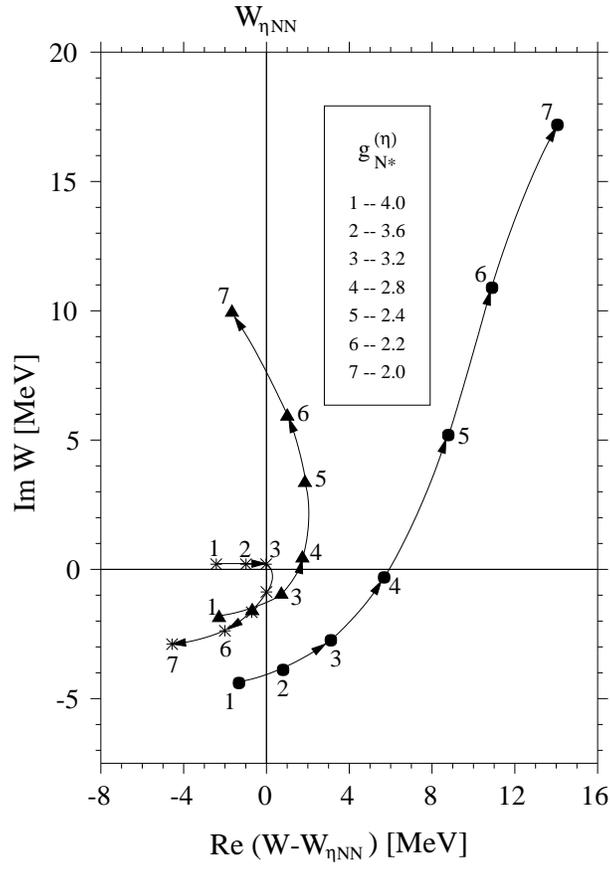,width=8cm,angle=0}}
\vspace{.5cm}
\caption{
The pole trajectories of the $\eta NN$ scattering matrix for the 
$(J^\pi;T)=(0^-;1)$ state. Notation as in Fig.~\protect{\ref{fig8}}.
}
\label{fig9}
\end{figure}

\begin{figure}
\centerline{\psfig{figure=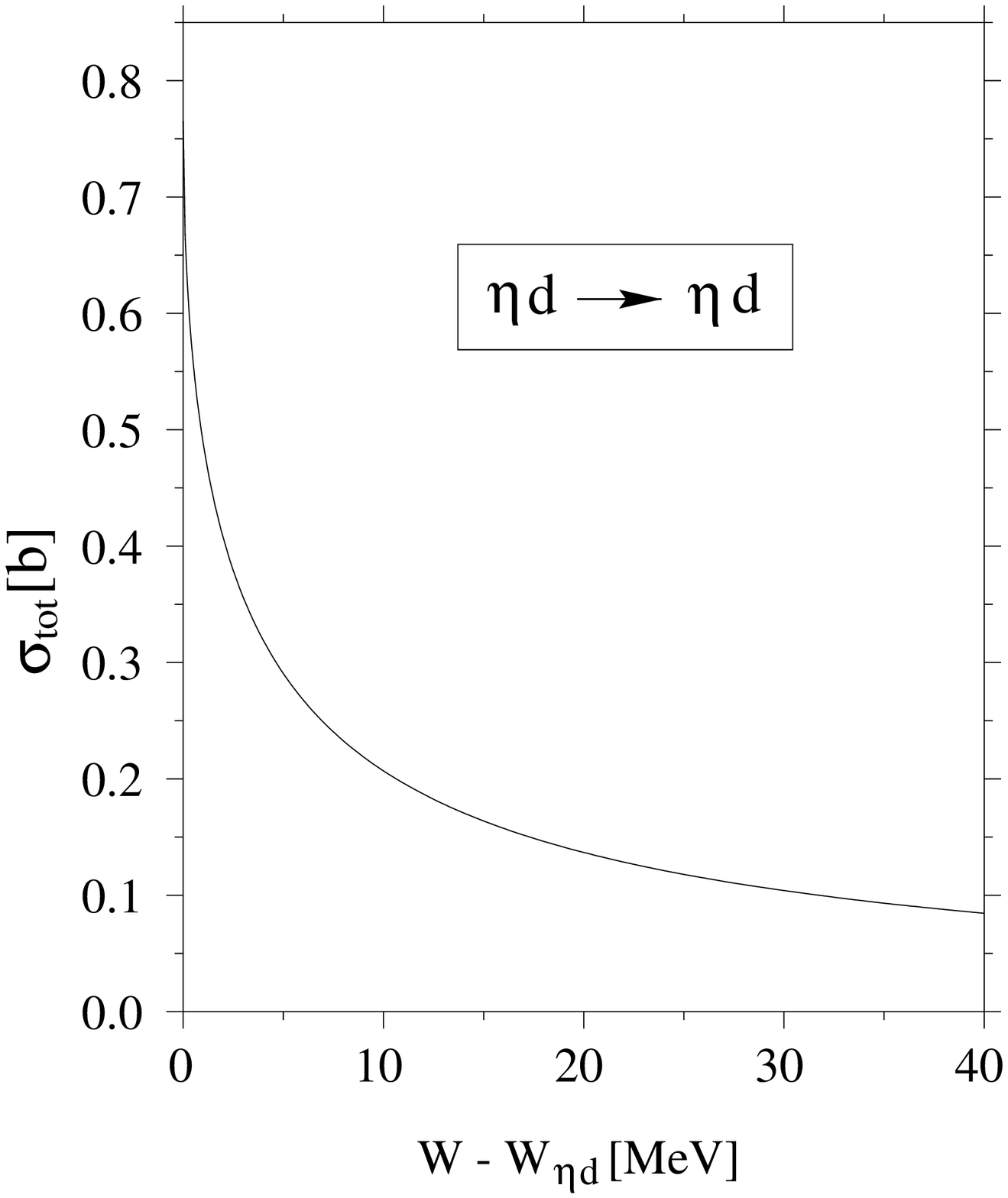,width=7cm,angle=0}}
\vspace{.5cm}
\caption{
Elastic $\eta d$ cross section evaluated according to the approximate
formula (\protect{\ref{sigED}}).
}
\label{fig10}
\end{figure}

\begin{thebibliography}{99}

\bibitem{Brookhaven}
R.E.\ Crien, et al., Phys.\ Rev.\ Lett.\ {\bf 60}, 2595 (1988)

\bibitem{Ueda}
T.\ Ueda, Phys.\ Rev.\ Lett.\ {\bf 66}, 297 (1991)

\bibitem{Tryas}
V.A.\ Tryasuchev, Phys.\ Atom.\ Nucl.\ {\bf 60}, 186 (1997)

\bibitem{Shev}
N.V.\ Shevchenko, V.B.\ Belyaev, S.A.\ Rakityansky, S.A.\ Sofianos,
W.\ Sandhas, nucl-th/9908035

\bibitem{Garc}
H.\ Garcilazo, M.T.\ Pe\~na, nucl-th/0002056

\bibitem{Rakit}
S.A.\ Rakityansky, S.A.\ Sofianos, V.B.\ Belyaev, W.\ Sandhas,
Phys.\ Rev.\ C {\bf 58}, R2043 (1998)

\bibitem{Wycech}
A.M.\ Green, S.\ Wycech, Phys.\ Rev.\ C {\bf 55}, R2167 (1997)

\bibitem{Wilk}
C.\ Wilkin, nucl-th/9810047

\bibitem{Metag}
V.\ Metag, in {\it Proceedings of the 8th International Conference on the 
Structure of Baryons, Bonn, 1998}, edited by D.M.\ Menze and B.\ Metsch
(World Scientific, Singapore 1999), p.\ 683

\bibitem{np}
H.\ C{\'a}len, et al., Phys.\ Rev.\ Lett.\ {\bf 80}, 2069 (1998)

\bibitem{BeTa}
C.\ Bennhold, H.\ Tanabe, Nucl.\ Phys.\ A {\bf 530}, 625 (1991)

\bibitem{FiTr}
V.A.\ Tryasuchev, A.I.\ Fiks, Phys.\ Atom.\ Nucl.\ {\bf 58}, 1168 (1995)

\bibitem{Abu}
L.J.\ Abu-Raddad, J.\ Piekarewicz, A.J.\ Sarty, M.\ Benmerrouche,
Phys.\ Rev.\ C {\bf 57}, 2053 (1998)

\bibitem{FiAr}
A.\ Fix, H.\ Arenh{\"o}vel, Z.\ Phys.\ A {\bf 359}, 427 (1997)

\bibitem{AGS}
E.O.\ Alt, P.\ Grassberger, W.\ Sandhas, Nucl.\ Phys.\ B {\bf 2}, 167 (1967)

\bibitem{Yam}
Y.\ Yamaguchi, Phys.\ Rev.\ {\bf 95}, 1628 (1954)

\bibitem{PDG98}
Particle Data Group, {\it Review of Particle Physics}, Eur.\ Phys.\ J.\ C 
{\bf 3}, 1 (1998)

\bibitem{Orlov}
K.\ M{\"o}ller, Yu.V.\ Orlov, Fiz.\ Elem.\ Chast.\ Atom.\ Yadra
{\bf 20}, 1341 (1989)

\bibitem{Glock}
W.\ Gl{\"o}ckle, Phys.\ Rev.\ C {\bf 18}, 564 (1978)

\bibitem{Moll}
K.\ M{\"o}ller, Czech.\ J.\ Phys.\ {\bf 32}, 291 (1982)

\bibitem{Kolg}
E.A.\ Kolganova, A.K.\ Motovilov, Phys.\ Atom.\ Nucl.\ {\bf 60}, 177 (1997),
nucl-th/9602001

\bibitem{Bel}
V.B.\ Belyaev, K.\ M{\"o}ller, Z.\ Phys.\ A {\bf 279}, 47 (1976)

\bibitem{Mats}
A.\ Matsuyama, K.\ Yazaki, Nucl.\ Phys.\ A {\bf 534}, 620 (1991)

\bibitem{Tay}
J.R.\ Taylor, {\it Scattering Theory} (John Wiley $\&$ Sons, New York 1972)

\bibitem{Land}
L.D.\ Landau, E.M.\ Lifshitz, {\it Quantum Mechanics,
Nonrelativistic Theory}
(Pergamon Press, Oxford, 1962)

\bibitem{Arenh}
H.\ Arenh{\"o}vel, Nucl.\ Phys.\ A {\bf 247}, 473 (1975)

\end{thebibliography}
\end{document}